\documentclass[aps,prl,groupedaddress]{revtex4-1}

\usepackage{graphicx}
\usepackage{xcolor}
\usepackage{dcolumn}
\usepackage{amsmath}
\usepackage{amsfonts}
\usepackage{bm}
\usepackage{epstopdf}
\usepackage{picture}
\usepackage{float}
\usepackage{dcolumn}
\usepackage{amsmath}
\usepackage{amssymb}
\usepackage{amsfonts}
\usepackage{bm}
\usepackage{picture}

\usepackage{caption}
\usepackage{booktabs} 
\usepackage{stmaryrd}
\usepackage[capitalise]{cleveref}
\usepackage{soul}
\usepackage{setspace}

\newcommand{\evec}{\mathbf{e}}

\newcommand{\Evec}{\mathbf{E}}

\newcommand{\mr}[1]{\mathrm{#1}}

\begin{document}

\title{Unidirectional Inter-Axial Coupling and Spontaneous Cooling in a~Non-Hermitian Dynamics of a~Levitated Particle}

\author{Tereza Zem\'{a}nkov\'{a}\textsuperscript{1,2}}
\author{Martin \v{S}arbort\textsuperscript{1}}
\author{Petr J\'{a}kl\textsuperscript{1}}
\author{Jan Ježek\textsuperscript{1}}
\author{Martin~\v{S}iler\textsuperscript{1}}
\author{Stephen H. Simpson\textsuperscript{1}}
\author{Pavel Zem\'{a}nek\textsuperscript{1}}
\author{Oto Brzobohat\'y\textsuperscript{1}}

\email{otobrzo@isibrno.cz}

\affiliation{\textsuperscript{1}The Czech Academy of Sciences, Institute of Scientific Instruments, Kr\'{a}lovopolsk\'{a} 147, 612 00 Brno, Czech Republic\\
\textsuperscript{2}Department of Optics, Palack\'{y} University, 17. listopadu 12, 771 46 Olomouc, Czech Republic}

\doublespacing

\begin{abstract}

\noindent Non-Hermitian dynamics in open systems can give rise to a variety of fascinating non-equilibrium phenomena, ranging from symmetry-breaking transitions to directional energy flow. Parity-time (PT) symmetry breaking determines the occurrence of dynamical instabilities, while non-reciprocal interactions enable asymmetric energy transfer between modes.
Here, we present a versatile optomechanical platform based on a vacuum-levitated nanoparticle that allows full control over the coupling of its mechanical modes, including non-reciprocal and non-conservative interactions. By engineering the spatial ellipticity and polarization of the trapping beam, we continuously tune the system from a reciprocal to a strongly non-reciprocal regime. This allows us to observe PT-symmetry phase transitions and to isolate a unidirectional regime in which one mode remains effectively decoupled while driving the other.
We demonstrate that elliptical polarisation of the trapping beam spanning unidirectional and reciprocal regimes induces asymmetric intermodal energy transfer. This results in the spontaneous cooling of one mechanical mode without external feedback.
Both modes share identical mass, size, charge, and optical environment, providing a clean and robust setting for exploring non-Hermitian dynamics, exceptional-point physics, and energy redistribution in minimal systems. Combined with recent advances in ground-state cooling, our results provide a direct route to realising non-Hermitian phenomena in the quantum regime.

\end{abstract}

\maketitle

\section{Introduction}
While many physical systems can be accurately described within the idealised framework of Hermitian dynamics, where reciprocal interactions ensure energy conservation and inherent stability, the increasing complexity of many-body systems has stimulated significant interest in non-Hermitian dynamics. These systems are characterised by non-reciprocal interactions or coupling to external reservoirs, as described in the literature \cite{ashida_nonhermitian_2020,  feng_nonhermitian_2017, ozturk_observation_2021, xu_topological_2016,regensburger_paritytime_2012,doppler_dynamically_2016, liska_ptlike_2024a, reisenbauer_nonhermitian_2024a, li_nonhermitian_2021}.
This field naturally encompasses open systems that exchange energy, momentum or information with their environment. A particularly rich subclass of non-Hermitian physics involves non-reciprocal interactions, which violate the intuitive symmetry of action and reaction. This leads to phenomena that are impossible to observe in equilibrium conditions.
Such interactions arise in diverse physical contexts, ranging from biological systems to photonics, electrical circuits, and mechanical metamaterials, enabling phenomena such as unidirectional transport, enhanced sensing, and robust mode control \cite{fruchart_nonreciprocal_2021, huber_topological_2016, mao_maxwell_2018,brandenbourger_nonreciprocal_2019}. 
Mechanical platforms with tunable non-reciprocity \cite{svak_transverse_2018, rieser_tunable_2022, reisenbauer_nonhermitian_2024a} offer a particularly transparent setting for exploring these effects.

Levitational optomechanics \cite{millen_quantum_2020,gonzalez-ballestero_levitodynamics_2021} provides a~uniquely versatile platform in this context. Focused laser beams can trap and manipulate mesoscopic particles under high vacuum, offering robust control over translational and rotational degrees of freedom. This level of control has enabled milestone achievements such as ground-state cooling of multiple motional modes \cite{delic_cooling_2020, piotrowski_simultaneous_2023, pontin_simultaneous_2023} and precision force sensing \cite{moore_searching_2021, geraci_sensing_2015,skrabulis_nanomechanical_2026a}, approaching levitated particles to the forefront of quantum optomechanics. Traditionally, these advances rely on the approximately conservative, potential-like nature of optical forces, which guarantees dynamical stability and enables thermodynamic control.
However, optical forces are fundamentally non-conservative \cite{roichman_influence_2008,wy_direct_2009,jiang_decomposition_2017,sukhov_nonconservative_2017, li_nonhermitian_2021}, but this feature becomes dominant 
when spatial or internal symmetries are broken—through particle anisotropy \cite{arita_coherent_2020a,arita_cooling_2023a, hu_structured_2023}, polarization structure \cite{svak_transverse_2018}, phase retardation \cite{rieser_tunable_2022, reisenbauer_nonhermitian_2024a, liska_ptlike_2024a}.
In such situations, the linearized local dynamics are governed by a generalized Hooke’s law with a non-symmetric stiffness matrix, leading to biased stochastic motion \cite{simpson_firstorder_2010}, limit-cycle oscillations \cite{simpson_stochastic_2021}, synchronization \cite{brzobohaty_synchronization_2023a}, and chiral dynamics. These effects naturally place levitated optomechanical systems within the broader framework of non-Hermitian physics, where gain, loss, and non-reciprocity play a central role.

Recent experiments have extended levitational control from single nanoparticles to arrays of optically interacting objects \cite{arita_optical_2018a, svak_stochastic_2021, rieser_tunable_2022, vijayan_cavitymediated_2024,reisenbauer_nonhermitian_2024a,liska_ptlike_2024a,liska_cold_2023}, demonstrating scalability and site-resolved readout. In such multi-particle systems, light-induced dipole–dipole interactions enable tunable coupling that can be continuously adjusted from reciprocal  to non-reciprocal and even unidirectional regimes. While these platforms provide rich collective dynamics, they inherently involve multiple particles with potentially non-identical masses, sizes, electric charges and trapping conditions, which can complicate precise control and interpretation of the interactions. 


Here, we adopt a complementary approach by exploiting a single levitated nanoparticle, whose motion along perpendicular axes 
serves as a model of coupled harmonic oscillators. This architecture offers intrinsic advantages: the oscillators share identical properties (mass, size, electric charge, refractive index), experience the same optical environment, but can be independently addressed and controlled via external force fields. As a result, the coupling between oscillators (interaxial coupling) can be engineered with exceptional precision and stability.
Using this platform, we demonstrate tunable coupling between the transverse motional modes of a single levitated mechanical oscillator, thereby realizing a controlled transition  from reciprocal to non-reciprocal and ultimately to fully unidirectional dynamics. Remarkably, within the non-reciprocal regime, an imbalance in the coupling strengths leads to asymmetric energy exchange between the two oscillators and, under special circumstances, leads to  
a spontaneous cooling of one oscillator 
accompanied by heating of the other. The effect arises from the interplay between non-conservative optical forces and viscous dissipation, and does not require external feedback or engineered reservoirs. The high degree of control and isolation provided by our system, therefore, enables a clean realization of tunable intermodal energy transport at the level of individual degrees of freedom. Our results establish levitated optomechanics as a~minimal and highly controllable platform for exploring non-Hermitian dynamics, directional energy flow, and autonomous thermalization phenomena.

\section{Results}
\subsection{Experimental platform}\label{sec2} 

\begin{figure*}[htb]%
    \centering
    \includegraphics[width=0.95\textwidth]{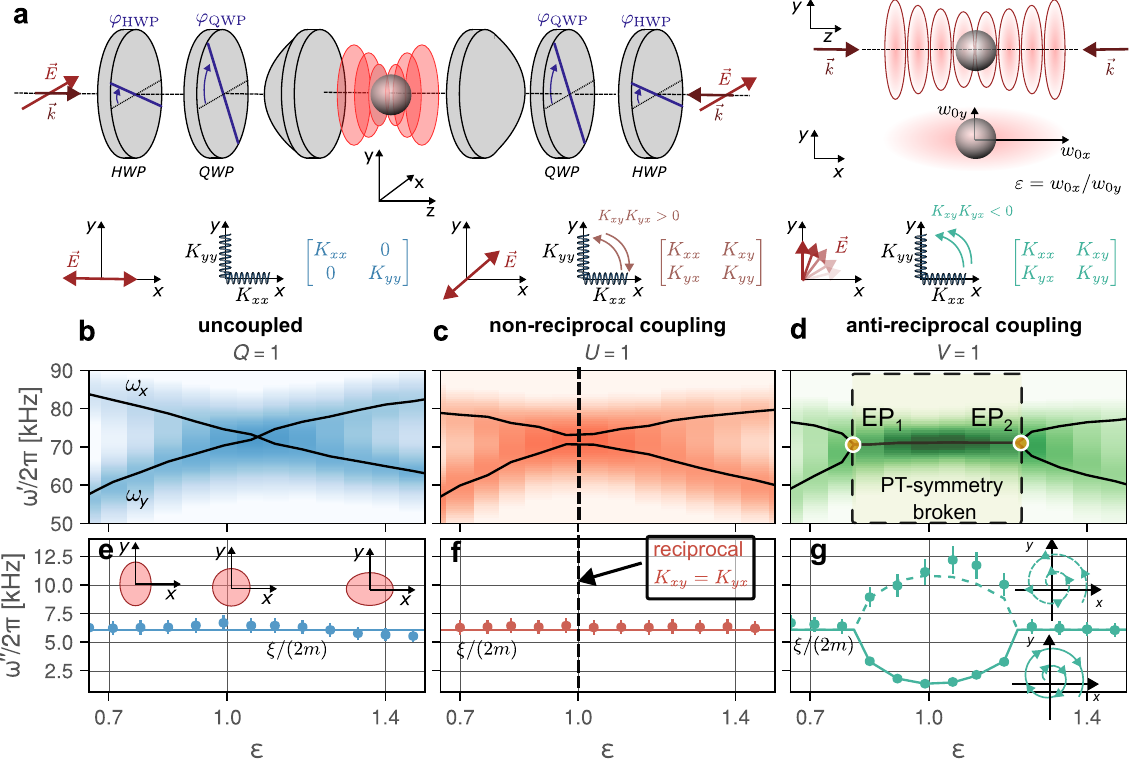} 
    \caption{\textbf{Overview of the experiment.}
\textbf{a,} Optical trap formed by counter-propagating (CP) Gaussian beams with tunable lateral elliptical intensity profile and arbitrary polarization, controlled using sets of wave plates (HWP -- Half-wave plate, QWP -- Quarter-wave plate). 
\textbf{b-d,} The sum of the power spectral densities (PSDs) of the particle motion along $x$ and $y$ axes is shown, together with the real parts of the eigenfrequencies 
for different values of the beam ellipticity $\varepsilon = w_{0x}/w_{0y}$.
\textbf{b,} Under horizontal linear polarization ($Q=1$), the particle motion along the $x$ and $y$ axes remains uncoupled. Which is illustrated by crossing of the eigenfrequencies $\omega_x$ and $\omega_y$, a behavior characteristic of independent oscillators  with a diagonal stiffness matrix.
\textbf{c,} Laser beam polarization is linear tilted by $\pi/4$ with respect to the horizontal axis, providing non-reciprocal (except $\varepsilon = 1$) inter-axial coupling defined by non-zero and generally non-equal off-diagonal terms $K_{xy}K_{yx} > 0$. 
The coupling is indicated experimentally by the avoided crossing of eigenfrequencies and stable dynamics with equal damping $\omega''$ at both eigenfrequencies given by the ambient gas.  
\textbf{d,} Circularly polarized trapping beam induces anti-reciprocal coupling, characterized by off-diagonal stiffness terms $K_{xy}K_{yx} < 0$. 
By varying the trap ellipticity $\varepsilon$, the system is driven through a pair of exceptional points (EPs) into a regime where the real parts of the eigenfrequencies coalesce. 
In this broken PT-symmetry phase, the imaginary parts bifurcate, representing the emergence of effective gain and loss (on top of the viscous dissipation due to the surrounding gas) see part \textbf{g}. 
\textbf{e, f,} 
Imaginary parts of the eigenfrequencies, reflecting viscous dissipation due to the surrounding gas 
$\xi/(2m)$. The complex eigenfrequencies are derived from stiffness matrices reconstructed from the measured particle trajectories. In the imaginary part, the discrete points represent results obtained by individually fitting the damping rate $\xi/(2m)$ for each ellipticity value $\varepsilon$. The solid curves depict the results from fitting assuming a constant damping rate (mean value obtained for horizontal polarization).
\label{fig1}}
\end{figure*}

Our experimental system is based on two counter-propagating (CP) Gaussian optical beams, forming a standing wave with an axial spacing between neighboring intensity minima of $\Delta z = \lambda/2=1064/2$\,nm. 
A silica microsphere with a diameter of $d = 987$\,nm is trapped near a node of this standing wave, where it is tightly confined along the axial direction.
The transverse ($x-y$) intensity distribution of the standing wave is generaly elliptic, controlled by the spatial light modulator (SLM), with axes oriented along $x$ and $y$ and thus characterized by beam waist radii $w_{0x}$ and $w_{0y}$.
The particle oscillates with larger amplitudes in the transverse $x$–$y$ plane, where the trap stiffnesses are lower, 
while oscillation amplitude along the $z$ axis is much lower due to higher axial trap stiffness, making this axial mode effectively decoupled from the lateral modes. 
As the SLM enables independent adjustment of the lateral beam waist radii, the trapping stiffnesses along the $x$ and $y$ axes can be tuned individually, see Fig.~\ref{fig1}a and Methods. 
Furthermore, by employing combinations of half- and quarter-wave plates, we precisely control the polarization state of the optical trap.
It determines the nature of the inter-axial force coupling, ranging from 
reciprocal to non-reciprocal, including the unidirectional regimes.

The stochastic motion of the mechanical oscillators with mass $m$ is governed by the linearized Langevin equation, which in the time domain reads
\begin{equation}
m \ddot{\mathbf{r}} = - \mathbf{K}\mathbf{r} - \mathbf{\Xi}\dot{\mathbf{r}} + \mathbf{F}^{\mathrm{L}}(t),
\label{eq:Langevin_general}
\end{equation}
where the lateral displacement vector is $\mathbf{r}(t) = [x(t), y(t)]^{\mathrm{T}}$, and the superscript $\mathrm{T}$ denotes vector transposition. The hydrodynamic drag is described by the diagonal matrix $\mathbf{\Xi}$, whose diagonal elements correspond to the ambient gas damping coefficient $\xi$.
The Langevin force $\mathbf{F^{\mathrm{L}}}(t)$ is uncorrelated in time with zero mean 
$\langle \mathbf{F^{\mathrm{L}}}(t) \rangle  = \mathbf{0},$  and variance $\langle \mathbf{F^{\mathrm{L}}}(t) \otimes  \mathbf{F^{\mathrm{L}}}(t')  \rangle  = 2 k_{\mathrm{B}} T  \mathbf{\Xi} \delta(t-t')$.
Here, $k_{\mathrm{B}}$ is the Boltzmann constant, $T$ is the thermodynamic temperature and the symbol $\otimes$ denotes the outer product of two column vectors.

For our geometry of the standing wave along $z$ axis, the electric field of the trapping beam in the transverse plane can be written as \cite{carter_electromagnetic_1972}
\begin{equation}
\Evec_{\rm{t}} = E_0 (\alpha \evec_{x} + \beta \mathrm{e}^{i\gamma} \evec_{y}) \exp \left[ -\frac{1}{2} \left( \frac{x^2}{w^2_{0x}} + \frac{y^2}{w^2_{0y}} \right) \right], 
\label{eq:pole}
\end{equation}
where the spatial profile is characterized by the beam waist radii $w_{0x}$ and $w_{0y}$, and the polarization state is defined by the normalized Stokes parameters 
\begin{align}
\begin{split}
 Q & = \alpha^2 - \beta^2,\\
 U & = 2\alpha\beta \cos(\gamma), \\
 V & = 2\alpha\beta \sin(\gamma), \\
 1 & = Q^2+U^2+V^2.
\end{split}    
\end{align}

For an arbitrary polarization state of the optical trap of an elliptical intensity profile, the stiffness matrix $\mathbf{K}$ can be expressed as a linear combination of stiffness matrices corresponding to the four principal polarization states: $\mathbf{K}^{(\mathcal{X})}$ and $\mathbf{K}^{(\mathcal{Y})}$ corresponding to the 
linear polarization along the $x$ and $y$ axes giving $Q=\pm1$; 
$\mathbf{K}^{(\mathcal{U})}$ for linear polarization tilted at $\pi/4$ with $U= 1$; and $\mathbf{K}^{(\mathcal{V})}$ for left-hand circular polarization with $V=1$
\begin{equation}
\mathbf{K} =
\begin{bmatrix}
K_{xx} & K_{xy} \\
K_{yx} & K_{yy}
\end{bmatrix} = 
 \begin{bmatrix}
\frac{1}{2}(1+Q)K^{(\mathcal{X})}_{xx}+\frac{1}{2}(1-Q)K^{(\mathcal{Y})}_{xx} & U K^{(\mathcal{U})}_{xy} + V K^{(\mathcal{V})}_{xy} \\
U K^{(\mathcal{U})}_{yx} + V K^{(\mathcal{V})}_{yx} & \frac{1}{2}(1+Q)K^{(\mathcal{X})}_{yy}+\frac{1}{2}(1-Q)K^{(\mathcal{Y})}_{yy}
\end{bmatrix}.
\label{eq:kmatrix}
\end{equation}
Based on the system's symmetries, the stiffness matrices $\mathbf{K}^{(\mathcal{X})}$ and $\mathbf{K}^{(\mathcal{Y})}$ are purely diagonal and oscillators are uncoupled. 
In contrast, the remaining matrices are generally non-Hermitian and characterize the non-reciprocal interactions. 
For a particle of the dimensions considered here, $\mathbf{K}^{(\mathcal{U})}$ and $\mathbf{K}^{(\mathcal{V})}$ possess unequal off-diagonal terms, $K_{xy} \neq K_{yx}$, fulfilling the conditions 
$K_{xy}^{(\mathcal{U})} K_{yx}^{(\mathcal{U})} > 0$ and $K_{xy}^{(\mathcal{V})} K_{yx}^{(\mathcal{V})} < 0$. 
These asymmetric off-diagonal elements represent the non-conservative optical forces that drive the inter-axial dynamics and enable directional energy transfer. 

The dynamics of the trapped particle are described by the complex eigenfrequencies of the normal modes derived from the general Langevin equation
\begin{equation}
    \omega_{\pm,n} = \pm \sqrt{ \frac{\Lambda_{n}}{m}  - \frac{\xi^2 }{4m^2}} + \mathrm{i} \frac{\xi}{2m} \equiv \pm \omega_{\pm,n}' + \mathrm{i} \omega_{\pm,n}'' ,
    \label{eq:frequency_omega_c_Lambda}
\end{equation}
where $n \in \{1, 2\}$ is the mode index. 
The eigenvalues of the stiffness matrix $\mathbf{K}$ are $\Lambda_{n} = \Lambda_{0} \pm \sqrt{D}$, 
with the mean stiffness $\Lambda_{0}$ and the discriminant $D$ defined as
\begin{align}
    \Lambda_{0} &= \frac{K_{xx} + K_{yy}}{2} , \label{eq:eigenvalues_K_Lambda0} \\
    D &= \frac{(K_{xx} - K_{yy})^2}{4} + K_{xy} K_{yx} . \label{eq:eigenvalues_K_LambdaD}
\end{align}

In the limit of weak inter-modal coupling ($\sqrt{|D|} \ll \Lambda_{0}$), the physical behavior is governed by the sign of the discriminant $D$. 
If $D > 0$, stiffness eigenvalues are real even if the interaction is non-reciprocal. The eigenfrequencies therefore have distinct real parts, corresponding to two oscillatory modes with different frequencies. Both modes experience the same viscous damping, so the imaginary parts of the eigenfrequencies are equal and the dynamics are stable  
\begin{equation}
    \omega_{n} \approx  (\omega_0 \pm \Delta \omega) + \mathrm{i} \frac{\xi}{2m} .
    \label{eq:frequency_omega_c_Lambda_real}
\end{equation}
Conversely, for $D < 0$, the system undergoes a transition to the regime where the stiffness eigenvalues form a complex conjugate pair, and the splitting $\Delta \omega$ shifts from real to the imaginary part of the eigenfrequencies, leading to emergence of two modes, one with increased damping by $+\Delta\omega$ (loss) and second with reduced damping by $-\Delta\omega$ (gain) which may results into unstable non-equilibrium dynamics 
\begin{equation}
    \omega_{n} \approx  \omega_0 + \mathrm{i} \left( \frac{\xi}{2m} \pm \Delta \omega \right) .
    \label{eq:frequency_omega_c_Lambda_imag}
\end{equation}
Here, $\omega_0 = \sqrt{\Lambda_0/m - \xi^2/4m^2}$ represents the fundamental trapping frequency and $\Delta \omega = \sqrt{|D|} / (2m\omega_0)$ characterizes the coupling-induced frequency shift. Within this framework, the total frequency splitting between the two modes is $2\Delta \omega$. At $D=0$ the system reaches an exceptional point, where both eigenvalues and eigenvectors coalesce. 

In the limit of negligible viscous damping, the system exhibits a spectral transition governed by the sign of the discriminant $D$. For $D > 0$, the stiffness eigenvalues remain real and the corresponding eigenfrequencies split symmetrically around the fundamental frequency $\omega_0\pm\Delta\omega$, yielding two stable oscillatory modes with purely real spectra. This regime is directly analogous to the unbroken phase of parity--time (PT) symmetric systems.

In contrast, for $D < 0$, the stiffness eigenvalues form a complex conjugate pair and the eigenfrequencies acquire identical real parts but opposite imaginary components $\omega_0\pm \mathrm{i}\Delta\omega$, resulting in one exponentially amplified mode and one damped mode which corresponds to the broken PT-symmetric phase. The transition point at $D = 0$ defines an exceptional point, where both eigenvalues and eigenvectors coalesce.

A closely related behavior has been extensively studied in classical PT-symmetric systems of coupled oscillators introduced by Bender \textit{et al.}~\cite{bender_systems_2014,bender_observation_2013}, where PT symmetry is enforced by a balanced distribution of gain and loss between two oscillators. 
In such systems, the phase transition between unbroken and broken PT symmetry is controlled by the coupling strength: below a critical value, the eigenfrequencies are real, while above it they bifurcate into complex conjugate pairs.

In the present system, however, gain and loss do not arise from explicitly balanced dissipative elements. Instead, they emerge effectively from the non-Hermitian structure of the stiffness matrix associated with non-conservative, non-reciprocal forces. 
Different normal modes couple to this force in qualitatively distinct ways: modes whose motion aligns with the force experience amplification, whereas modes opposing it are damped. As a result, the system reproduces the same spectral transition as a PT-symmetric system, despite lacking an explicit gain--loss balance.

This distinction highlights that, unlike canonical PT-symmetric systems, where symmetry breaking is driven by tuning the balance between gain/loss and coupling, the transition observed here is governed by the interplay between modal structure, their mechanical frequencies, and non-reciprocal coupling. The resulting behavior is therefore more generally understood as a phase transition with PT-symmetric spectral characteristics, rather than a strict realization of PT symmetry.


\begin{figure*}[htb]%
    \centering
    \includegraphics[width=0.85\textwidth]{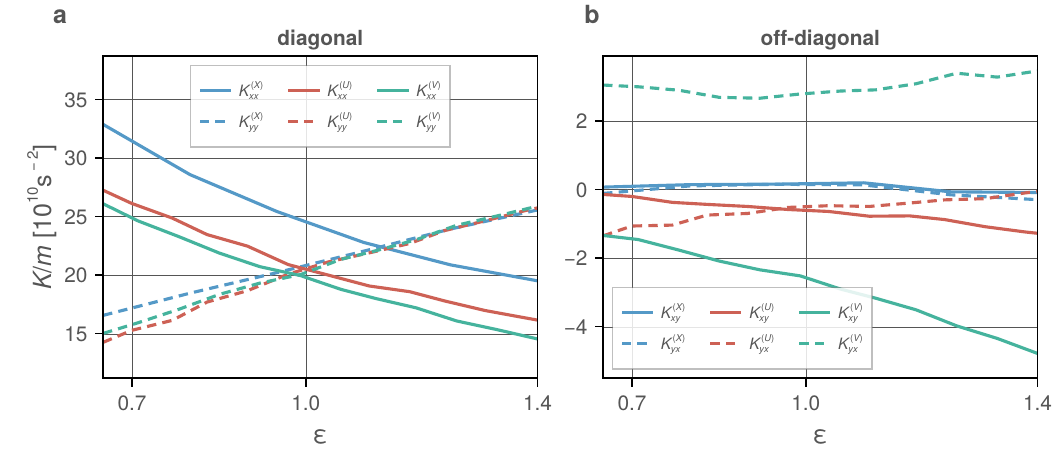}
    \caption{\textbf{Experimentally determined stiffness matrices.}
Diagonal (\textbf{a}) and off-diagonal (\textbf{b}) elements of the mass-weighted stiffness matrices $\mathbf{K}/m$ for the three polarization configurations studied in Fig.~\ref{fig1}. The blue curves correspond to horizontal linear polarization ($Q=1$), for which the off-diagonal elements vanish across the entire range of beam ellipticity. The red curves represent linear polarization rotated by $\pi/4$ ($U=1$), while the green curves correspond to circular polarization ($V=1$). These configurations exhibit non-conservative interactions that fulfill the theoretically predicted conditions: $K_{xy}^{(\mathcal{U})}K_{yx}^{(\mathcal{U})} > 0$ for stable non-conservative coupling, and $K_{xy}^{(\mathcal{V})}K_{yx}^{(\mathcal{V})} < 0$ for unstable coupling, respectively. For a symmetric spatial intensity profile ($\varepsilon = 1$), the system obeys the expected symmetry relations: $K_{xy}^{(\mathcal{X})} = K_{yx}^{(\mathcal{X})} = 0$  for linear horizontal polarization, $K_{xy}^{(\mathcal{U})} = K_{yx}^{(\mathcal{U})}$ for linear $\pi/4$ polarization, and $K_{xy}^{(\mathcal{V})} = -K_{yx}^{(\mathcal{V})}$ for circular polarization. The relative uncertainties (not shown) of the estimated stiffness matrix components are approximately 1\%. 
}
    \label{fig2}
\end{figure*}

To characterize the tunability of the oscillation frequencies $\omega_{\pm,n}'$ arising from the elliptical profile of the optical trap, in Fig.~\ref{fig1}b we first consider linear polarization along the horizontal $x$-axis ($Q=1$). 
In this configuration, the stiffness matrix defined by Eq.~(\ref{eq:kmatrix}) is diagonal, and the two oscillators are uncoupled ($K_{xy} = K_{yx} = 0$) in the linear regime. 
Figure~\ref{fig1}b shows the experimentally measured sum of the power spectral densities (PSDs) for $x$ and $y$ particle motion 
as a function of the trap ellipticity defined as $\varepsilon = w_{0x}/w_{0y}$. 
By keeping the geometric mean waist $w_{\mathrm{av}} = \sqrt{w_{0x}w_{0y}} = 1.65$\,\textmu m constant, the laser power remains also approximately constant throughout the experiment. 
By varying $\varepsilon$, we tune the oscillation frequencies across the 60–80\,kHz range. The observed spectra exhibit a pure frequency crossing, which is a characteristic of two independent, uncoupled oscillators.
In the same configuration, Fig.~\ref{fig1}e provides the imaginary parts of the eigenfrequencies, $\omega_{\pm,n}^{''}$, which are identical and given solely by the damping rate $\xi/(2m)$ from surrounding gas, ensuring stable particle dynamics.

\begin{table*}[ht]
\caption{Summary of coupling regimes, stiffness matrix symmetry for varying polarization and trap ellipticity $\varepsilon$ ($\varepsilon_\mathrm{EP1}$ and $\varepsilon_\mathrm{EP2}$ denote exceptional points, see Fig.~\ref{fig1}d). }
\centering
\footnotesize
\begin{tabular}{llllll}
\toprule
\textbf{Polarization} & \textbf{Trap Ellipticity} & \textbf{Stiffness Matrix} & \textbf{{System Properties}} & \textbf{Eigenfrequencies } & \\ \midrule
\hline 
$Q=1$ & Any $\varepsilon$ & $K_{xy}=K_{yx}=0$ &  stable &  $ \omega_{x,y}'$, & $\omega_{x,y}'' = \xi/2m$ \\
linear $x$ & $D\geq0$ & & uncoupled & &  \\ 
\hline \\
$U = 1$ & $\varepsilon=1$ & $K_{xy}=K_{yx}\neq 0$ & PT-symmetric, stable & $ \omega_{x,y}' \approx \omega_0 \pm \Delta\omega$, & $\omega_{x,y}'' = \xi/2m$ \\
linear $\pi/4$ & $D\geq0$ & & reciprocal & &  \\ 
& $\varepsilon \neq 1$ & $K_{xy} \neq K_{yx} $ and & PT-symmetric, stable &  $ \omega_{x,y}' \approx \omega_0 \pm \Delta\omega$, & $\omega_{x,y}'' = \xi/2m$ \\
& $D\geq0$ & $K_{xy}K_{yx} > 0$ & non-reciprocal & &  \\ 
\hline \\
$V=1$ & $\varepsilon \leq  \varepsilon_\mathrm{EP1}$ or $\varepsilon_\mathrm{EP2} \leq \varepsilon$ & $K_{xy}K_{yx} <0$ & PT-symmetric, stable & $ \omega_{x,y}' \approx \omega_0 \pm \Delta\omega$, & $\omega_{x,y}'' = \xi/2m$\\
 circular & $D\geq0$ &  & anti-reciprocal & &  \\ 
  & $\varepsilon_\mathrm{EP1} < \varepsilon < \varepsilon_\mathrm{EP2}$& $K_{xy}K_{yx} < 0$ & PT-symmetry broken, unstable &  $ \omega_{x}' = \omega_{y}' = \omega_0$, & $\omega_{x,y}'' \approx \xi/2m \pm \Delta\omega$ \\
    & $D<0$&  & anti-reciprocal &  &  \\
\bottomrule
\end{tabular}

\label{tab:coupling_regimes}
\end{table*}

Next, by rotating the linear polarization to $\pi/4$ ($U=1$), we achieve conservative coupling between the oscillators only for a symmetric beam profile ($\varepsilon = 1$). 
In this case, the off-diagonal stiffness elements are identical $K_{xy}^{(\mathcal{U})} = K_{yx}^{(\mathcal{U})}$, defining a system with fully reciprocal inter-axial coupling (see Fig.~\ref{fig2}b).
In contrast, for non-symmetric beam profiles, these elements become unequal $K_{xy}^{(\mathcal{U})} \neq K_{yx}^{(\mathcal{U})}$ resulting in a non-reciprocal 
 inter-axial coupling. 
It results in a non-conservative yet stable system across the entire range of beam ellipticities because the product $K_{xy}^{(\mathcal{U})}K_{yx}^{(\mathcal{U})} > 0$ ensures positive definiteness of the stiffness matrix (i.e. $D>0$) and the imaginary part of the eigenfrequencies is defined solely by the gas damping independent on beam ellipticity $\varepsilon$ (see Fig.~\ref{fig1}f).
An avoided crossing of the eigenfrequencies demonstrates the coupling in the spectral domain (see Fig.~\ref{fig1}c). 
However, the coupling strength, typically characterized by the minimal frequency difference $2\Delta \omega / \omega_0 = 0.03$ 
is relatively weak here due to wider beam waists of our optical trap
compared to other platforms~\cite{liska_ptlike_2024a,reisenbauer_nonhermitian_2024a}.

Finally, by employing circular polarization ($V=1$), the spin-induced optical force introduces an azimuthal component (see Fig.~\ref{fig1}d), engineering an anti-reciprocal coupling between the $x$ and $y$ axes~\cite{svak_transverse_2018,antognozzi_direct_2016,li_nonhermitian_2024}.
In this configuration, the off-diagonal elements of the stiffness matrix are of  opposite signs ($K_{xy}K_{yx} < 0$). 
Consequently, the system can undergo a transition from a PT-symmetric phase—characterized by oscillations at two distinct real eigenfrequencies $\omega_{n}'$ with a common damping rate $\omega_{n}'' = \xi/(2m)$—to a broken PT-symmetry phase. In the broken phase, the real parts of the eigenfrequencies coalesce to a single value, while the imaginary parts bifurcate into two distinct damping rates, $\omega_{n}'' \approx \xi/(2m) \pm \Delta \omega$ (see Eq.~\ref{eq:frequency_omega_c_Lambda_imag} and Fig.~\ref{fig1}g).

For the oscillation mode with increased damping, where $\omega_{+}'' \approx \xi/(2m) + \Delta\omega$, the dynamics of mode remain stable. Here, the non-conservative optical force opposes the particle motion, effectively dissipating kinetic energy and causing the trajectories to spiral toward the fixed point at $x = y = 0$ (dashed curve, Fig.~\ref{fig1}g). 
Conversely, for the mode with reduced damping, $\omega_{-}'' \approx \xi/(2m) - \Delta \omega$, the non-conservative force injects energy into the system. 
This leads to an outward spiral that drives the particle away from equilibrium (solid curve, Fig.~\ref{fig1}g). 
This PT-symmetry broken region is bounded by exceptional points (EPs), which occur when  $D = 0$. Table~\ref{tab:coupling_regimes} summarizes the key features of the coupling regimes discussed. Notably, the stable reciprocal inter-axial coupling, often assumed in standard optical trapping, represents only a specific case within a broader landscape of dynamical regimes.

\begin{figure*}[htb]%
    \centering
    \includegraphics[width=0.95\textwidth]{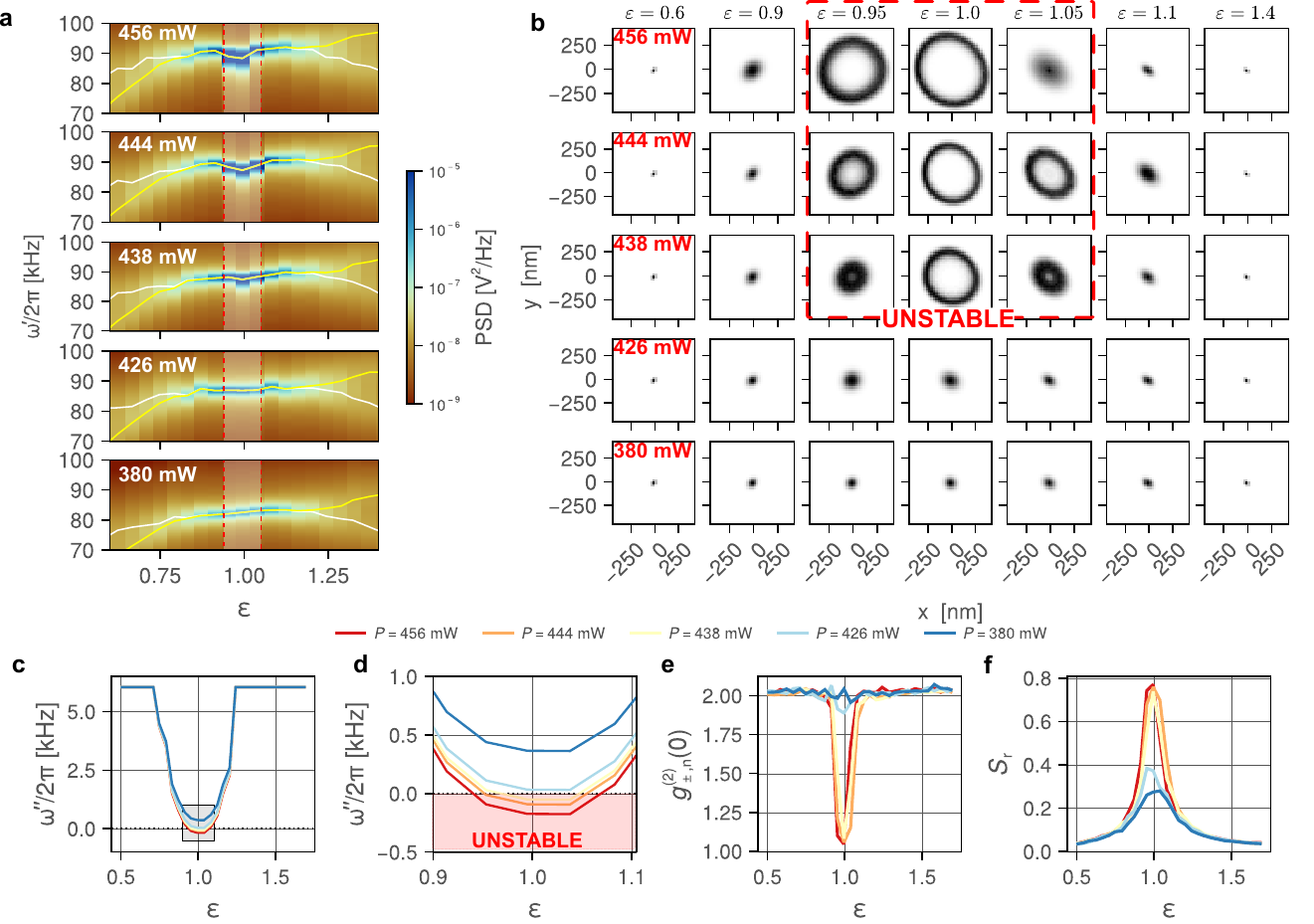}
    \caption{\textbf{Stability above threshold.}  
\textbf{a,} Summed power spectral density (PSD) depicting oscillation amplitude growth with increasing optical power.  
\textbf{b,} Spatial $xy$ probability density functions illustrating the transition from a stable fix point dynamics to a limit cycle for ellipticities within the range $0.95 \leq \varepsilon \leq 1.05$ as the power surpasses the stability threshold. 
\textbf{c, d,} Eigenfrequency analysis revealing the crossing of the imaginary components $\omega_{-,n}''$ from positive to negative values, i.e., surpassing the stability threshold and confirming the dynamical instability. 
\textbf{e,} Evaluation of statistical properties of instantaneous local particle energy in stable regime, where the second-order coherence function $g^{(2)}(0) \approx 2$ (thermal state), and in limit cycles at $P > 430$~mW, where $g^{(2)}(0)$ approaches unity (coherent state).
\textbf{f,} Relative Shannon entropy $S_r$  as a function of system parameters, quantifying phase-space localization and the emergence of synchronized motion within the limit-cycle regime.}
\label{fig3}
\end{figure*}

Figure~\ref{fig2} summarizes the experimentally reconstructed diagonal (panel a) and off-diagonal (panel b) elements of the stiffness matrix for the three polarization configurations discussed above. The stiffness matrices are obtained by fitting analytically derived time-dependent covariance functions -- constructed from position and velocity coordinates  -- to the corresponding covariance functions calculated from the experimental time traces (see Methods and Supplementary Information for details).
The simultaneous reconstruction of the stiffness matrix elements and the effective viscosity allows us to determine the eigenfrequencies of the system and thereby fully characterize the oscillators dynamics and their mutual interaction. Both the diagonal and off-diagonal stiffness elements show good qualitative agreement with the predictions of our numerical model (not presented). Small deviations arise from experimental imperfections, which result in a residual non-zero value of the off-diagonal elements for horizontal polarization, $Q=1$, although these terms are expected to vanish for all values of the trapping-beam ellipticity.

\subsection{Non-linear anti-reciprocal interaction}

For circular polarization, when we reduced viscosity $\xi/(2m)$ or increased optical power  effective damping rate becomes negative ($\omega_{-}^{''} < 0$), the amplitude of the amplified eigenmode increases exponentially $a_{-} \propto \exp(\mathrm{i} \omega_{-}^{'}t) \exp(- \omega_{-}^{''}t)$ and the system undergoes a Hopf bifurcation and transits to dynamical instability, here the linear theory breaks down. 
In this case, the particle motion starts exploring the intrinsic nonlinearity in the optical trap which acts to stabilize the oscillation amplitude, which results in either limit-cycle formation or eventually particle loss from the trap \cite{svak_transverse_2018, liska_ptlike_2024a,reisenbauer_nonhermitian_2024a}. 
The evolution of the system dynamics as a function of the trapping laser power (ranging from 380 to 456 mW) is presented in Fig.~\ref{fig3}. 
The  summed PSD of $x$ and $y$ particle oscillations reveals a significant amplitude increase of the amplified mode with increasing optical power in the unstable 
region (see Fig.~\ref{fig3}a). 
Above the instability threshold \cite{svak_transverse_2018} ($P > 430$\,mW), a clear shift of the oscillation frequency to the lower values emerges for trapping beam ellipticity close to $\varepsilon=1$. 
Increasing oscillation amplitudes reach the region of 
the weakening Duffing nonlinearity in the Gaussian trapping beam where the oscillation frequency decreases with the square of the oscillation amplitude.
Spatial dynamics are further illustrated by the $xy$ probability density function (Fig.~\ref{fig3}b), which confirms the formation of the limit cycle for trapping beam  ellipticities having value between 0.95 and 1.05 once the power surpasses the bifurcation threshold. 
This transition is verified by eigenfrequency analysis (Figs.~\ref{fig3}c,d), where the imaginary part of eigenfrequency $\omega_{-}^{''}$ shifts from positive to negative values when the laser power increases. 
In this analysis, the stiffness matrix is reconstructed in the stable fixed-point regime ($P_0=380$\,mW), where the linearized theoretical model is valid. 
The resulting stiffness elements are then rescaled to  this power to account for different trapping laser powers $\mathbf{K}(P) = P/P_0\mathbf{K}(P_0)$ and  consequently complex eigenfrequencies $\omega_{n}^{''}$ are calculated.

In the above-threshold regime, the limit-cycle oscillator exhibits statistical characteristics typical of a phonon laser~\cite{liska_ptlike_2024a,kuang_nonlinear_2023,pettit_optical_2019,zheng_arbitrary_2023,vahala_phonon_2009,sharma_$mathcalpt$_2022}. 
This state is identified by a well-defined threshold condition, a characteristic spectral linewidth narrowing, and coherent oscillations. 
These dynamics are quantified by the second-order coherence function at zero time delay, $g_{-}^{(2)}(0) = \langle E_{-}^2 \rangle / \langle E_{-} \rangle^2$, where $E_-$ represents the instantaneous total energy. 
Within the instability region, $g_{-}^{(2)}(0)$ approaches unity, a typical value for a coherent state. 
This is in stark contrast to the value of 2 observed in the below-threshold regime (Fig.~\ref{fig3}e), which corresponds to the statistics of a thermal state.
The mean value $\langle E_{-} \rangle$ carries information about the limit cycle amplitude increase. 
Finally, the emergence of synchronized motion of both oscillators within the limit cycle regime is quantified using relative Shannon entropy $S_r$ ~\cite{tass_detection_1998, brzobohaty_synchronization_2023a}.  In this context, $S_r$ measures the strength of phase locking between motions of the coupled oscillators, taking values between zero and one, where a value of one indicates perfect phase synchronization. For the observed limit-cycle oscillations, we find $S_r \approx 0.8$, indicating near-perfect phase locking between the $x$ and $y$ degrees of freedom (see Fig.~\ref{fig3}f).

Finally, we note that the ability to continuously control the trapping-beam ellipticity provides a practical route to stabilizing spinning  non-spherical or anisotropic particles in circularly polarized optical vacuum traps \cite{arita_cooling_2023a}. The restored stability induced by the ellipticity of the optical trap would enable rapid, stable spinning of anisotropic particles of intermediate size, such as micro-rods of birefringent micro-spheres.

\begin{figure*}[htb]%
    \centering
    \includegraphics[width=0.95\textwidth]{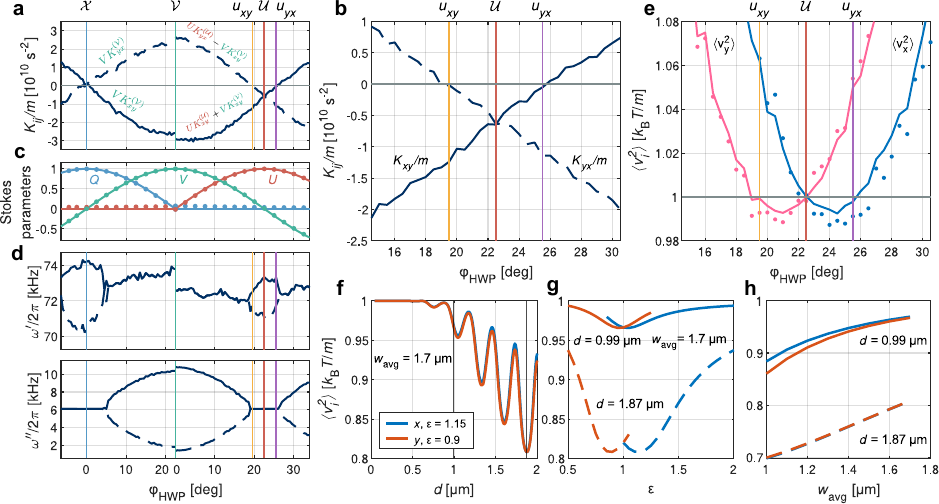}
    \caption{\textbf{Experimental scan of polarization for a beam with symmetric intensity profile $\varepsilon=1$.} \textbf{a,b} Off-diagonal elements of the stiffness matrices for scans from horizontal linear polarization $Q=1$ to circular polarization $V=1$ and following to 
        linear polarization $U=1$.   
        \textbf{c,} Experimental (dots) and theoretical values of normalized Stokes coefficients $(Q,V,U)$ describing the corresponding polarization of the trapping beam. 
        \textbf{d,} The real and imaginary parts of the eigenfrequencies calculated from stiffness matrices reconstructed from experimental particle trajectory.  Colored vertical lines indicate the polarization states $\mathcal{X}$, $\mathcal{V}$, and $\mathcal{U}$, as well as the unidirectional coupling regimes $u_{xy}$ and $u_{yx}$.
        \textbf{e, } Normalized mean kinetic energy $\langle v^2 \rangle / (k_\mr{B}T/m)$ along the $x$ and $y$ axes in the vicinity of unidirectional coupling conditions ($K_{yx}=0$, yellow; $K_{xy}=0$, violet).  
        A reduction in kinetic energy below the equipartition limit is observed in the regions $K_{xy} < K_{yx} < 0$ and $K_{yx} < K_{xy} < 0$, respectively, illustrating a spontaneous cooling originated from non-reciprocal energy transfer. Full curves denote values determined using reconstructed stiffness matrix and dots denote the values computed directly from the experimental velocity coordinates. 
     Plots \textbf{f,} \textbf{g,}\textbf{h,} analyze the value of the minimal kinetic energy as a function of particle diameter $d$, beam ellipticity $\varepsilon$ and 
    the average beam waist $w_\mathrm{avg}$, respectively, from numerical calculations of stiffness matrices for corresponding parameters.}
    \label{fig4}
\end{figure*}

\subsection{Unidirectional interaction and spontaneous cooling}

As demonstrated in Fig.~\ref{fig2}b, our system is almost perfectly symmetric for a circular trapping beam  $\varepsilon = 1$. In this case, the off-diagonal elements of the stiffness matrix exhibit well-defined symmetry properties: for horizontal linear polarization, they vanish, $K_{xy}^{(\mathcal{X})}~=~K_{yx}^{(\mathcal{X})}~=~0$; for linear polarization oriented to $\pi/4$, they are equal and non-zero, $K_{xy}^{(\mathcal{U})} = K_{yx}^{(\mathcal{U})} \neq 0$; and for circular polarization, they are antisymmetric, $K_{xy}^{(\mathcal{V})} = -K_{yx}^{(\mathcal{V})}$.
In the following, we investigate the continuous tunability of the optical interaction between the two oscillators, represented by particle motion along $x$ and $y$ axes, by varying the polarization state of the trapping beam (see Fig.~\ref{fig4}). 
All measurements were performed for a fixed beam ellipticity $\varepsilon = 1$ and consist of two consecutive polarization scans between the basis polarization states.

In the first scan, the polarization was varied from horizontal linear polarization ($Q=1$) to circular polarization ($V=1$), see values of normalized Stokes coefficients measured for right-hand side trapping beam in Fig.~\ref{fig4}c.
This was achieved by rotating the half-wave plates while keeping the quarter-wave plates fixed at $\varphi_{\mathrm{QWP}} = 0$. 
In this configuration, the off-diagonal stiffness elements evolve from a non-interacting regime, $K_{xy}^{(\mathcal{X})} = K_{yx}^{(\mathcal{X})} =  0$, to pure non-reciprocal and non-conservative interaction characterized by $VK_{yx}^{(\mathcal{V})}$ (and analogously for the other off-diagonal element), see Figs.~\ref{fig4}a and  \ref{fig4}b.
When the half-wave plate (HWP) is oriented at approximately $5^\circ$, the system undergoes a phase transition -- from a PT-symmetric phase with distinct eigenfrequencies to a PT-symmetry-broken phase. 
In this broken phase, the particle oscillates along $x$ and $y$ axes with the same frequency $\omega'$ but decays with two different damping rates given by $\omega''$, 
as given by Eq.~(\ref{eq:frequency_omega_c_Lambda}), Table~\ref{tab:coupling_regimes} and demonstrated in Fig.~\ref{fig4}d. 
Notably, the imaginary component of these eigenfrequencies reaches extremal values for pure circular polarization ($V = 1$), as illustrated in Fig.~\ref{fig4}c.

More interestingly, the second scan -- starting from circular polarization ($V = 1$), passing through linear polarization at $\pi/4$ ($U = 1$), and continuing to circular polarization of opposite handedness ($V = -1$) -- reveals a continuous evolution of the system. The dynamics progress from a non-reciprocal regime ($K_{xy} K_{yx} < 0$), through an exceptional point, into a unidirectional coupling regime ($K_{yx} = 0$), followed by a reciprocal configuration at $U = 1$. 
Upon further polarization tuning, the system enters a second unidirectional regime ($K_{xy} = 0$), crosses another exceptional point, and finally returns to a non-reciprocal regime ($K_{xy} K_{yx} < 0$) (see Fig.~\ref{fig4}a–\ref{fig4}d).
The second scan was performed experimentally by rotating the half-wave plates with the quarter-wave plates fixed at $\varphi_{\mathrm{QWP}} = \pi/4$. 
For a specific range of elliptical polarizations (highlighted by yellow and purple lines and magnified in Fig.~\ref{fig4}b), the system enters a unidirectional coupling regime in which motion along one axis drives the orthogonal axis without reciprocal back-action. This regime is characterized by the stiffness matrix in which only one off-diagonal element, $U K_{xy}^{(\mathcal{U})} + V K_{xy}^{(\mathcal{V})}$ (or, equivalently, $U K_{yx}^{(\mathcal{U})} + V K_{yx}^{(\mathcal{V})}$), is non-zero, while the other vanishes. 
This behavior arises from the distinct symmetry properties of the two polarization states: for $U$ polarization, the off-diagonal elements always have the same sign, $K_{xy}^{(\mathcal{U})} K_{yx}^{(\mathcal{U})} > 0$, whereas for $V$ polarization they always have opposite signs, $K_{xy}^{(\mathcal{V})} K_{yx}^{(\mathcal{V})} < 0$. Consequently, by choosing an appropriate superposition of $U$ and $V$ polarizations, one can cancel one of the off-diagonal stiffness elements while retaining the other. 

In the region between reciprocal ($U=1$) and unidirectional ($u_{xy}$) interactions, where $K_{xy} < K_{yx} < 0$  (Fig.~\ref{fig4}b) the oscillator along $x$ axis experiences net energy gain, whereas the other (along $y$ axis) is effectively cooled. 
This effect can be visualized (see solid lines in Fig.~\ref{fig4}e) using the analytically derived normalized variance of velocities, which is directly related to the kinetic energy and temperature of the centre of mass of the oscillators (see Methods and Supplementary Information)
\begin{equation}
\begin{aligned}
\langle v_{x}^2 \rangle &= \frac{k_B T}{m} \left[1 + \frac{K_{xy} (K_{xy} - K_{yx})}{ 2 (K_{xy} K_{yx} + K_{xx} \frac{\xi^2}{m})}\right], \\
\langle v_{y}^2 \rangle &= \frac{k_B T}{m} \left[1 + \frac{K_{yx} (K_{yx} - K_{xy})}{ 2 (K_{xy} K_{yx} + K_{xx} \frac{\xi^2}{m})}\right].
\end{aligned}
\label{eq:varv}
\end{equation}
We compare these values with normalized velocity variances obtained numerically from
particle trajectories (dots); both approaches consistently show an increase of energy along one axis, $\langle v_x^2 \rangle/\left(\frac{k_B T}{m}\right) > 1$, and a reduction along the orthogonal axis, $\langle v_y^2 \rangle/\left(\frac{k_B T}{m}\right) < 1$, indicating heating of the $x$ mode and spontaneous cooling of the $y$ mode relative to thermal equilibrium.
Conversely, when the inequality is reversed, $K_{yx} < K_{xy} < 0$, the roles of the two oscillators are interchanged, with the $y$ mode gaining energy and the $x$ mode being cooled.
This phenomenon is associated with non-reciprocal interaxial interactions and energy transfer between the oscillators, which modifies the equipartition theory for non-Hermitian systems  \cite{li_nonhermitian_2024}.

Figures~\ref{fig4}f–\ref{fig4}h present a theoretical parameter sweep across experimentally relevant conditions to identify the optimal regime for spontaneous cooling. We analyze the influence of particle size, beam ellipticity, and average beam waist, showing that the cooling effect is enhanced for larger particles and more tightly focused optical traps. The normalized kinetic energy is evaluated using stiffness matrix elements obtained from numerical calculations, combined with analytical expressions for the velocity variances (see Methods and Supplementary Information for details).

\section{Discussion}

In this work, we have established a platform for engineering non-reciprocal 
interactions between transverse motional modes of a single levitated nanoparticle. Rather than focusing on the observation of specific coupling regimes, we consider here how the ability to continuously tune both the spatial and polarization degrees of freedom of the trapping field addresses a central challenge in non-Hermitian physics: the controlled realization of symmetry breaking and directional coupling within an open system of two mechanical oscillators, realized by a single levitating particle oscillating along $x$ and $y$ axes, coupled by light. 

Compared to existing implementations of non-Hermitian systems—such as coupled optical resonators, electronic circuits, or multi--particle optomechanical arrays--our single-particle approach minimizes parameter inhomogeneity and environmental asymmetries. This intrinsic uniformity allows the stiffness matrix to be interpreted directly in terms of fundamental force gradients, thereby providing a clearer connection between microscopic optical fields and emergent non-Hermitian dynamics. At the same time, this simplicity imposes constraints: the accessible coupling strengths and dissipation channels are ultimately limited by the optical trapping configuration, which may restrict exploration of strongly nonlinear or many-body regimes.

Methodologically, reconstructing the stiffness matrix from experimental data provides a direct, model-independent way to quantify non-reciprocal interactions. However, this approach assumes linear response and weak coupling between modes, and its extension to regimes with strong fluctuations or nonlinearities remains an open question. Future work could address these limitations by incorporating feedback control or by extending the analysis to time-dependent modulation of the trapping field.

The emergence of directional energy flow occurs in the intermediate non-reciprocal regime between reciprocal and fully unidirectional coupling, where asymmetric inter-modal interactions enable net energy transfer between the oscillators. This places our system within the broader framework of non-Hermitian transport and exceptional-point physics, where directionality and sensitivity enhancements play a central role. In contrast to conventional gain–loss engineered systems, where non-Hermiticity arises from explicit dissipation or amplification, our approach realizes non-reciprocal dynamics through controlled spatial and polarization structuring of the optical field, providing an alternative route to non-Hermitian behavior.

The ability to engineer such interactions in a levitated platform opens opportunities for exploring non-Hermitian phase transitions in low-dissipation environments and for interfacing with quantum optomechanical protocols. In particular, extending these concepts to the quantum regime could enable new strategies for state control, sensing, and the realization of synthetic mechanical matter with programmable non-reciprocal interactions.

\clearpage
\section{Materials and Methods}
\subsection{Experimental setup}

We employed two counter-propagating laser beams to create the optical trap. By applying different phase masks to the spatial light modulator (SLM), the ellipticity of the laser beam could be tuned from circular to elliptical. In addition, the total trapping power was controlled by adjusting the diffraction efficiency of the first and higher diffraction orders. A combination of a half-wave plate and a quarter-wave plate allowed precise control of the polarization state in the optical trap, enabling systematic transitions between linear and circular polarization. In the case of circularly polarized trapping beams, the polarization may have either of two opposite handedness. A detailed schematic of the experimental setup is shown in Fig.~S1 in the Supplementary Information.

For optical trapping, silica (SiO$_2$) particles (Microparticles, mean diameter 987\,nm) were dispersed in isopropyl alcohol and sonicated for approximately 30 minutes in an ultrasonic bath. The resulting suspension was then loaded into a nebulizer (Nimo HNK-MESH-01), which sprayed droplets containing the particles into the trapping region inside the vacuum chamber at atmospheric pressure.

\subsection{Particle position detection}

A quadrant photodiode (QPD, Hamamatsu Photonics G6849) was used to record the particle position in the $x$–$y$ plane. A custom-coated mirror (CM) reflecting 5\,\% of the trapping beam (for both $s$ and $p$ polarizations) directed a portion of the light toward the detector, as illustrated in Fig. S1. The QPD detection principle relies on homodyne scheme where we monitor the intensity distribution of the interference pattern formed by the trapping light and the light scattered by the particle. The resulting voltage signal is proportional to the particle position. The sampling frequency was set to 1\,MHz, and data segments of 500\,k samples (0.5\,s) were recorded.

An independent measurement of the particle position was obtained using a high-speed CMOS camera (i-SPEED 508). The particle was illuminated by a separate laser beam (Coherent Prometheus, wavelength 532\,nm) propagating perpendicular to the trapping beams, as shown in Fig. S1. This configuration enabled detection of the particle motion in the $y$–$z$ plane. A large beam waist ($w_0 = 40$\,\textmu m) and low optical power (approximately 0.4\,mW) minimized the influence of the illumination beam on the optical forces acting on the particle. The exposure time was set between 1\,\textmu s and 1.5\,\textmu s, with a sampling frequency ranging from 500\,kHz to 750\,kHz.

Offline particle tracking from the high-speed video recordings was performed by analyzing the symmetries of the particle images \cite{leite_threedimensional_2018}. This method provided measurements of the in-plane $y$ and $z$ coordinates with nanometer-scale precision.

\subsection{Adjusting polarization}

The polarization state of each trapping beam was controlled using a combination of a half-wave plate (HWP) and a quarter-wave plate (QWP), arranged according to the scheme shown in Fig.~\ref{fig1}. To verify the polarization state within the optical trap, the polarization was characterized using a polarimeter (Thorlabs PAX1000IR2/M) positioned downstream of the quarter-wave plate and both trapping lenses.

The initial alignment procedure was performed using only the half-wave plates, with the quarter-wave plates temporarily removed from the optical path. The HWPs were mounted on motorized rotation stages (Thorlabs K10CR2) with an absolute angular accuracy of $\pm0.14^{\circ}$. Using the polarimeter, the HWP angle $\varphi_\mathrm{HWP}$ was adjusted to obtain optimal horizontal linear polarization. The corresponding angular settings were recorded and used as the reference configuration for horizontal linear polarization.

After establishing this baseline, the quarter-wave plates were inserted into the optical path. Their orientations were manually adjusted to two specific angles: $\varphi_\mathrm{QWP}=\pi/4$, corresponding to circular polarization (right-handed for the left beam and left-handed for the right beam), and $\varphi_\mathrm{QWP}=0$, corresponding to horizontal linear polarization. The manual adjustment of the QWPs had an estimated precision of approximately $1^{\circ}$.

For each of these two QWP orientations, a detailed polarization scan of the trapping beam was performed by rotating the HWP over the range $0^{\circ}$ to $45^{\circ}$. The normalized Stokes parameters measured by the polarimeter were then compared with theoretical predictions, as shown in Fig.~\ref{fig4}c.

\subsection{Stiffness matrix from particle trajectories}

The stiffness matrix of the optical trap was determined directly from experimentally measured trajectories of the levitated particle in the $x$–$y$ plane. For each experimental configuration, the particle position was recorded as voltage signals $x'(t)$ and $y'(t)$ using a quadrant photodiode (QPD). The time series were band-pass filtered in the frequency range 10–150\,kHz in order to isolate the mechanical motion of the particle from low-frequency drift and high-frequency electronic noise. The corresponding velocity signals $v'_x(t)$ and $v'_y(t)$ were obtained by transforming the position signals into the frequency domain, multiplying by $\mathrm{i}\omega$, and applying the inverse Fourier transform. This procedure avoids numerical differentiation in the time domain and provides a stable estimate of the particle velocities.

To characterize the dynamics of the system, time-dependent covariance functions were constructed from all combinations of the particle position and velocity coordinates. Specifically, we evaluated the position–position, position–velocity, velocity–position and velocity–velocity covariance functions as functions of the time lag $\tau$, defined as $\langle r_i(t) r_j(t+\tau)\rangle$, $\langle r_i(t) v_j(t+\tau)\rangle$, $\langle v_i(t) r_j(t+\tau)\rangle$ and $\langle v_i(t) v_j(t+\tau)\rangle$, where $i,j \in \{x,y\}$. In total, sixteen time-dependent covariance functions were obtained from the experimental data. 
To reduce differences in magnitude between the various coordinate combinations and to improve numerical stability during fitting, we switched to normalized coordinates and correlation functions. The experimental signals were normalized by their standard deviations. The resulting dimensionless coordinates were then used to compute normalized correlation functions for time lags up to $\tau = 0.1$\,ms. 


The analytical expressions for the time-dependent covariance functions were derived from the linearized Langevin equation describing the stochastic motion of a particle confined in an optical trap characterized by a generally non-symmetric stiffness matrix $\mathbf{K}$ and viscous damping. In this framework the particle dynamics are governed by the interplay between the restoring optical forces, Stokes drag, and thermal Langevin noise. Solving the Langevin equation in the frequency domain yields expressions for the power spectral densities of the particle coordinates, which can be transformed into time-dependent covariance functions using the Wiener–Khinchin theorem. The resulting analytical covariances depend on the stiffness matrix elements $K_{ij}$, the viscous damping coefficient $\xi$, and the particle mass $m$, and can be expressed in terms of the complex eigenfrequencies of the normal modes of the system. The corresponding correlation functions were finally obtained by normalizing the coordinates using their respective standard deviations.

The analytical correlation functions were fitted simultaneously to the experimentally calculated correlations using nonlinear regression. All correlation functions share the same set of physical parameters and were therefore optimized together in a global fit. The fitting parameters included the stiffness matrix elements per unit mass, $\hat{K}_{ij}=K_{ij}/m$, the damping rate $\Gamma=\xi/m$, and calibration coefficients relating the measured voltage signals to the physical particle displacements along $x$ and $y$ directions. Initial estimates of the diagonal stiffness matrix elements were obtained from the peak frequencies of the power spectral densities of the particle motion. The particle mass was determined from the nominal particle size and material density, and the temperature was fixed to $T=300$\,K.

From the fitted parameters we reconstructed the full stiffness matrix of the optical trap and determined the corresponding eigenvalues and eigenfrequencies, which characterize the coupled mechanical modes of the system. Further details of the theoretical derivation of the covariance functions, the numerical implementation of the fitting procedure, and the full analytical expressions used in the analysis are provided in the Supplementary Information.

\subsection{Stiffness matrix modeling}
Equation (\ref{eq:pole}) describes the transversal components of the electric field intensity in the focal plane of the laser beam with general elliptical profile of intensity. 
The axial field component $E_z$ as well as the variation of the field along the propagation axis can be obtained using angular spectral decomposition \cite{carter_electromagnetic_1972}.
We should point out here that Eq.~(\ref{eq:pole}) does not apply for arbitrary small waists $w_{0x}$, and $w_{0y}$. 
In principle, we should have $w_{0x} \gg \lambda/(2\pi)$ and  $w_{0y} \gg \lambda/(2\pi)$. 
Otherwise, the exact shape of Eq.~(\ref{eq:pole}) could not be obtained without the evanescent components of the angular spectrum. 

The optical forces acting on a spherical particle located near a focus of a standing wave can be calculated by means of $T$-matrix method \cite{gouesbet_generalized_2011}.
In order to accelerate the optical force calculations the double integrals over the angular spectrum  can be simplified to a single integration and series sum using the Jacobi-Anger expansion for modified Bessel functions. 
This technique may also be employed to obtain beam shape coefficients describing the trapping field  in the base of vector spherical wave functions which enable fast calculations of optical forces \cite{gouesbet_generalized_2011}.
The calculation of optical forces in the standing wave are performed by employing mirror symmetry of the beam shape coefficients. 

Finally, the stiffness matrix $\mathbf{K}$ is calculated in 2 steps. 
Firstly, the axial stable position of the particle is numerically found by modifying the particle center $z$ coordinate. 
Secondly, the  $\mathbf{K}^{(i)}$ matrices are evaluated by numerical differentiations for all 4 principal polarization states ($\mathcal{X}, \mathcal{Y}, \mathcal{U}, \mathcal{V}$).
This procedure is then repeated for desired particle radius and beam waists; polarization state is included by Eq.~(\ref{eq:kmatrix}).

\section{Acknowledgement}
The Czech Science Foundation (GA23-06224S);  Akademie v\v{e}d \v{C}esk\'{e} republiky (Praemium Academiae); 
Ministerstvo \v{S}kolstv\'{i} ml\'{a}de\v{z}e a t\v{e}lov\'{y}chovy ($\mathrm{CZ.02.01.01/00/22\_008/0004649}$).


%

\end{document}